# The lubricity of mucin solutions is robust toward changes in physiological conditions

*Jian Song, Benjamin Winkeljann, Oliver Lieleg* *

*Department of Mechanical Engineering and Munich School of Bioengineering,*
*Technical University of Munich, Garching 85748, Germany*

\* Corresponding author:

Prof. Dr. Oliver Lieleg

Department of Mechanical Engineering and Munich School of Bioengineering,

Technical University of Munich,

Boltzmannstraße 11, 85748 Garching, Germany

E-mail: oliver.lieleg@tum.de

Phone: +49 89 289 10952; Fax: + 49 89 289 10801

ORCID iD:

0000-0002-2924-6640 (JS); 0000-0002-6334-6696 (BW); 0000-0002-6874-7456 (OL)

**Abstract：**

Solutions of manually purified gastric mucins have been shown to be promising lubricants for biomedical purposes, where they can efficiently reduce friction and wear. However, so far, such mucin solutions have been mostly tested in specific tribological settings, i.e. in combination with different material pairings; variations in the composition of the lubricating fluid have not been systematically explored yet. We here fill this gap and determine the viscosity, adsorption behavior, and lubricity of porcine gastric mucin solutions on hydrophobic surfaces at different pH levels, mucin and salt concentrations and in the presence of other proteins. We demonstrate that mucin solutions provide excellent lubricity even at very low concentrations of 0.01 % (w/v), over a broad range of pH levels and even at elevated ionic strength. Furthermore, we provide mechanistic insights into mucin lubricity, which help explain how certain variations in physiologically relevant parameters can limit the lubricating potential of mucin solutions. Our results motivate that solutions of manually purified mucin solutions can be powerful biomedical lubricants, e.g. serving as artificial joint fluids for viscosupplementation, as eye drops or mouth spray, or as a personal lubricant for intercourse.

# 1 Introduction

Many body fluids such as tears, saliva or the synovia act as aqueous lubricants, i.e., they play an essential role in reducing friction between rubbing tissue contacts. In the absence of sufficient lubrication, severe problems can develop which are typically associated with tissue inflammation, patient discomfort and pain. Examples include dry irritated eyes, vaginal dryness, dryness of the mouth impeding proper speech and mastication, and osteoarthritis, i.e. excessive wear formation on articulating cartilage surfaces [1-3].

To prevent those issues, artificial (bio)polymer based lubricants have been introduced to reduce friction and wear in biomedical settings [4, 5]. One particular macromolecule has recently gained lots of attention in this regard: mucin, the structural key component of mucus. Mucus is a slimy substance that is widely secreted by many organisms, and mucus protects the lung, eye, gastrointestinal tract, vagina, and other mucosal surfaces from pathogenic invaders, dehydration and mechanical damage [6]. Mucus hydrogels comprise water (> 90 wt.%), inorganic salts, mucins, and some minor components which depend on the mucus source [7]. Mucin glycoproteins have a high molecular weight (0.5–20 MDa) and are densely glycosylated.

Due to their outstanding performance as biolubricants, the tribological properties of mucin solutions have been widely studied during the past years. For instance, Pult *et al.*[8] investigated friction between the cornea and the eyelid as well as between the contact lens surface and the eyelid. Their results indicated that the tribological behavior of these interfaces correlates with the quantity and quality of the mucins in the tear film. Winkeljann *et al.*[9] demonstrated that purified gastric mucins can prevent the formation of tissue damage on porcine cornea – both as a lubricant solution and contact lens coating. Furthermore, mucins have been shown to prevent surface damage on articular cartilage [10] and the corneal epithelium [11].

Different from other lubricants, mucin-based solutions also provide excellent lubricity in the boundary lubrication regime. Depending on the material pairing tested and the tribological setup used, the measured friction coefficients $\mu$ range from 0.2 to less than 0.01 [12-15]. This excellent lubricity of mucin solutions is closely linked to the fact that mucins can adsorb to a wide variety of surfaces. As a consequence of mucin adsorption, hydrophobic (artificial and biological) surfaces are rendered hydrophilic, which facilitates the formation of a lubricating water film on those surfaces. This, in turn, leads to a separation of the load-bearing surfaces under shear. Moreover, cyclic shear-off and readsorption of mucins (which is referred to as sacrificial layer formation) and hydration lubrication, i.e., the generation of a surface-bound hydration layer provided by adsorbed mucins, further reduce friction [16]. In part, certain molecular motifs on the mucin glycoprotein have been identified which contribute to the mentioned lubrication mechanisms, and there are also first attempts to create synthetic, mucin-inspired lubricants [17-20].

Although the studies mentioned above are good examples of how recent research has gathered new insight into the molecular reasons that provide lubricity to mucin solutions, we are probably still far from appreciating the full range of applications for which mucin-based lubricants could be useful. In part, this is due to the limited availability of highly functional, lab-purified mucins. In contrast, as shown in **Fig. S1**, commercial, industrially purified mucins only show vastly inferior lubricity [9]. Hence, to further explore the lubricating potential of mucin-based solutions, it is crucial to use manually purified mucins which have maintained their outstanding function. In the past, this limitation might also have prevented a more detailed examination of biomedical applications where mucin-based solutions could be used as biolubricants. However, with this problem having received more attention and the purification process of porcine gastric mucins being developed further [21], this field is ready to be revisited.

Another important aspect which has not received enough attention yet is how components of a complex system such as the human body interact with mucin-based lubricants. Body fluids in general and mucosal tissues in particular can significantly differ in terms of pH, ionic strength and protein content, and each of those parameters might impact the performance of mucin-based lubricants. Thus, a systematic evaluation of those physiological parameters is needed to assess the potential of mucin solutions for biomedical applications, i.e. to use them as artificial joint fluid, eye drops, personal lubricant, or mouth spray (**Fig. 1**).

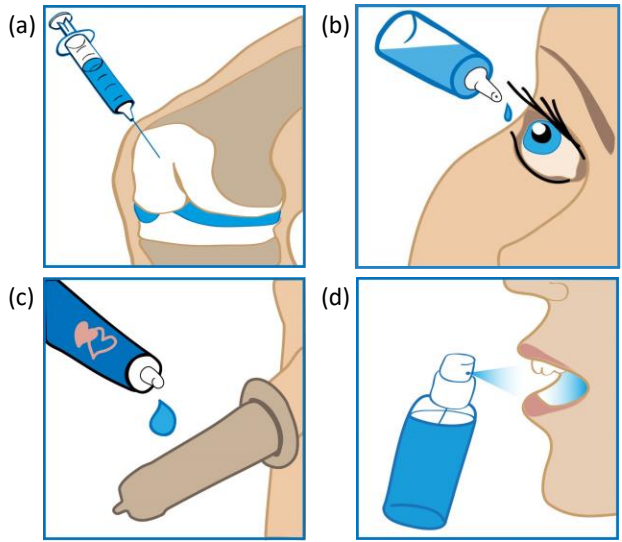

**Fig. 1** Potential use of mucin-based solutions in biomedical applications: (a) artificial joint fluid for viscosupplementation, (b) eye drops, (c) personal lubricant for intercourse, and (d) mouth spray.

The work presented here is motivated by such considerations. We evaluate the lubricity of solutions reconstituted from manually purified gastric mucins on hydrophobic surfaces and demonstrate that this lubricity is well-maintained when alterations in pH, mucin concentration or ionic strength are induced. Moreover, we describe how other solubilized proteins can interfere with mucin lubricity by competing for surface adsorption. In addition to exploring the operating range of mucin-based lubricants, our results also provide a better understanding of microscopic regulation mechanisms which affect the lubricating potential of mucin solutions.

# 2 Materials and methods

## 2.1 Mucin purification

Mucin purification was conducted as described earlier in detail [21, 22]. In brief, fresh porcine stomachs were opened, and food debris washed off with tap water. Then, crude mucus was collected by scraping the mucosal surface of the tissue. The obtained mucus was pooled and diluted in PBS buffer (10 mM, pH = 7.4) for overnight solubilization. Cellular debris and lipid contaminants were removed from this solubilized mucus via two centrifugation steps (first at 8300 x g at 4 °C for 30 min, then at 15000 x g at 4 °C for 45 min) and a final ultracentrifugation step (150000 x g at 4 °C for 1 h). Afterwards, the mucins were separated from other macromolecules by size exclusion chromatography using an Äkta purifier system (GE Healthcare, Chicago, IL, USA) and an XK50/100 column packed with Sepharose 6FF. Then, the mucin fractions were pooled, dialyzed against ultrapure water, and concentrated by cross-flow filtration. Finally, the concentrated mucins lyophilized and stored at -80 °C. All purified mucins were exposed to UV-light for 1 h for sterilization before use.

## 2.2 Buffer solutions

To avoid introducing artefacts associated with buffering substances, four kinds of buffer solutions with different components were tested. Phosphate buffer was prepared by dissolving selected amounts of sodium dihydrogen phosphate monohydrate (Sigma, St. Louis, USA) in ultrapure water and pH was adjusted to 7 by adding 32 % NaOH dropwise. HEPES buffer was prepared by dissolving 20 mM 4-(2-hydroxyethyl)-1-piperazineethanesulfonic acid (HEPES, Roth, Karlsruhe, Germany) in Millipore water and pH was adjusted to 7 by adding 0.1 M NaOH. Universal buffer (UB, pH 2.0-8.2) [23] and Britton-Robinson buffer (BRB, pH 1.9-11) [24] were prepared as described in the literature. In detail, UB was prepared by combining 20 mM HEPES, 20 mM 2-(N-morpholino)ethanesulfonic acid (MES, Sigma, St. Louis, USA) and 20 mM sodium acetate (Roth, Karlsruhe, Germany) in ultrapure water, and pH was adjusted with 10 M NaOH or 5 M HCl as needed. BRB was obtained by mixing 40 mM acetic acid (Roth, Karlsruhe, Germany), orthophosphoric acid (Roth, Karlsruhe, Germany) and boric acid (Roth, Karlsruhe, Germany) into ultrapure water, and the pH was adjusted with 0.2 M NaOH.

## 2.3 Lubricant solutions

Solutions with different mucin concentrations were generated in 20 mM HEPES, pH 7. Unless stated otherwise, a mucin concentration of 0.1 % (w/v) (= 1 mg/mL) was chosen to ensure comparability to earlier studies [16, 25]. To investigate the performance of mucin solutions at different pH, the mucins were dissolved in UB at pH 2, 4, 6 and 8, respectively. BSA (FraktionV T844.1, Roth), amylase (10065, Sigma), and lysozyme (L6876, Sigma) were dissolved in 20 mM HEPES buffer (pH 7) at a concentration of 80 mg/L, 300 U/mL and 26 mg/mL, respectively. Then, the protein solutions were mixed in a ratio of 1:1 with a 0.2 % (w/v) mucin solution, which was prepared as described above. Thus, final protein concentrations of 40 mg/mL BSA, 150 U/mL amylase, and 13 mg/mL lysozyme were achieved, which is similar to the concentration of those proteins in body fluids [26, 27]. To study the influence of the buffer ionic strength, different concentrations of NaCl were added to the mucin solution (in 20 mM HEPES buffer, pH 7) when preparing the test fluid.

## 2.4 Viscosity measurements

The viscosities of the mucin-based lubricant solutions used in this study were determined on a commercial shear rheometer (MCR 302, Anton Paar, Graz, Austria) using a cone-plate geometry (CP50-1, Anton Paar). For each measurement, 570 µL of the test solution was pipetted onto the stationary plate to fully fill the space between the measuring head and the sample plate. Measurements were conducted at 21 °C, and the shear rate was varied from 10 to 1000 $s^{-1}$. The viscosity values shown in table 1 for each solution represent averaged measurement results acquired at a shear rate of 100 $s^{-1}$ from three independent samples.

## 2.5 Tribological tests

A steel-on-PDMS tribo-pairing was chosen to evaluate the lubricity of mucin solutions. Steel spheres with a diameter of 12.7 mm (Kugel Pompel, Vienna, Austria) were used as received. PDMS pins were prepared as cylinders with a diameter of 6.1 mm. In detail, PDMS prepolymer and cross-linker (Sylgard 184, Dow Corning, Wiesbaden, Germany) were mixed in a ratio of 10:1. Then, the mixture was placed into a vacuum chamber for 1 hour to remove air bubbles. Afterwards, the solution was poured into a steel mold and cured at 80 °C for 4 h. Both, the steel spheres and PDMS pins were used without further polishing as they showed low roughness ($Sq_{steel}$ < 200 nm, $Sq_{PDMS}$ < 50 nm) when investigated with a laser scanning microscope (VK-X1100, Keyence, Osaka, Japan).

The tribological experiments were performed at 21 °C using the tribology unit (T-PTD 200, Anton Paar) of a commercial shear rheometer (MCR 302, Anton Paar) as described before [28]. In brief, three PDMS pins were mounted into a pin holder and washed with ethanol and ultrapure water. Then, 600 µL of a lubricant solution was applied to ensure full coverage of the pins. The normal load was chosen to be 6 N, resulting in an average contact pressure of ~0.3 MPa. The sliding velocity was varied from $10^{-5}$ to $10^{0}$ m/s to probe as many lubrication regimes as possible. For each condition, three independent experiments were carried out using a fresh set of PDMS pins for each measurement.

## 2.6 Adsorption measurements

The adsorption properties of mucins on hydrophobic PDMS surfaces were studied by quartz crystal microbalance with dissipation monitoring (QCM-D) using a qcell T-Q2 platform (3T-Analytik, Tuttlingen, Germany). Gold sensor chips were coated with a thin PDMS film to obtain similar hydrophobic surfaces as used in the tribological tests. To obtain this coating, PDMS prepolymer and cross-linker (Sylgard 184, Dow Corning, Wiesbaden, Germany) were mixed in a ratio of 10:1, and were further diluted with n-hexane to obtain a 1% (v/v) polymer solution. Then, a bare gold sensor chip was placed into the center of a spin coater (WS-400B-6NPP/LITE, Laurell, North Wales, USA), and 100 µL of the prepared PDMS mixture was pipetted onto the gold chip. To distribute the PDMS solution, the spin-coater was set into rotation - first at 1500 rpm for 20 s and then at 3000 rpm for 60 s. Afterwards, the coated sensor chip was cured at 80 °C for 4 h.

At the beginning of each adsorption test, a pure buffer solution (without any proteins, mucins or salts) was injected at a flow rate of 100 µL/min until a stable baseline was obtained. Afterwards, a mucin-based solution was injected at 100 µL/min for ~30 min to obtain an adsorption curve. The resulting frequency shift ($\Delta f$, Hz) and dissipation shift ($\Delta D$) were automatically calculated by the software "qGraph" (3T-Analytik, Tuttlingen, Germany). For each condition, three independent experiments were carried out. Two additional experiments were conducted if the observed difference between different conditions was small and a statistical analysis was necessary. Then, the obtained results were statistically evaluated by a One-way ANOVA (ANalysis Of VAriance) followed by a post-hoc Tukey HSD (Honestly Significant Difference) using the software "SPSS" (v20.0.0, SPSS Inc., Chicago, USA). For *p*-values smaller than 0.05, differences were considered to be statistically significant.

# 3 Results and discussion

To evaluate the potential of mucin solutions as biolubricants for human use, several physiological parameters need to be considered, which might affect mucin lubricity. We here focus on variations in pH and ionic strength as those two parameters can be quite different on different body sites. Moreover, we aim at identifying the minimal mucin concentration that still conveys good lubricity, and we ask if prominent examples of proteins from body fluids interfere with the lubricating properties of mucin. As a material pairing for the tribological experiments, we choose a steel-on-PDMS setup. This particular material pairing is selected, since both, steel and PDMS-based materials, are commonly used in various medical devices [29, 30], and such a hard-on-soft pairing involving both a hydrophilic and a hydrophobic surface is also frequently employed in biotribological studies [31, 32] to mimic, e.g., the tongue-palate interface. Furthermore, with this steel-PDMS paring, all three lubrication regimes are clearly visible, which – compared to other tribo-parings – makes it easier to analyze the tribological performance of different lubricants in detail (**Fig. S1**).

## 3.1 Choosing the right buffer system

When working with solutions of biological molecules such as mucins, using buffers to control pH and ionic strength is critical, especially if the influence of either parameter on properties of the solution is investigated. Although phosphate based buffers are frequently used in mucin tribology [33-35], one needs to be aware, that – depending on the material pairing studied – buffer substances may have a significant influence on the experimental outcome. For a steel/PDMS pairing, which is not only used in this study but is regularly employed in the field of biotribology [16, 19, 36-39], this can be indeed an issue: When the phosphate concentration in a standard PBS buffer is increased from 10 mM to 500 mM, the friction coefficient in the boundary lubrication regime drops by almost one order of magnitude (**Fig. 2a**). This can be attributed to the ability of phosphate ions to readily react with the steel surface – this is a frequently reported mechanism to modify the surface of steel components for industrial applications [40, 41]. Of course, phosphate concentrations as high as 500 mM are not physiologically relevant; however, we already detect a perceivable influence of phosphate ions on the tribological response of the system at the lowest concentration tested, i.e. 10 mM. Although such a phosphate buffered tribology system would still allow for direct comparisons between buffered biopolymer solutions and buffer only, it still disturbs the measurement outcome in two ways: First, one main advantage of a steel/PDMS pairing is, that all three lubrication regimes are clearly distinguishable; thus, the influence of different (macro)molecular ingredients can be easily identified. However, when using a phosphate-based buffer, the Stribeck curve is less pronounced and the maximal ambitus between 'poor lubricity' and 'great lubricity' is reduced. This makes it harder to detect improvements in lubricity provided by lubricating molecules. Second, a reaction of phosphate ions with the steel surface might cause side effects arising from a changed surface structure or surface chemistry, and this can complicate the interpretation of the results.

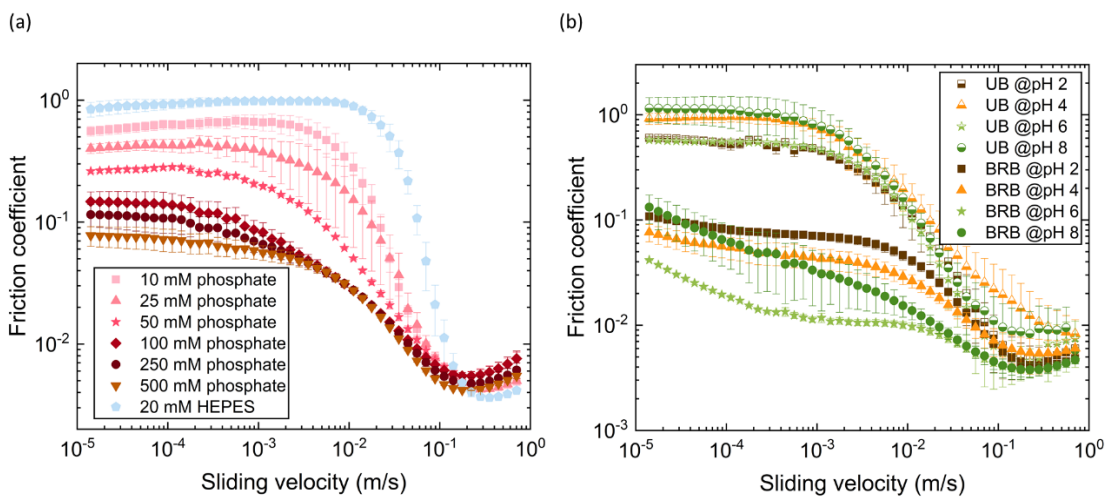

**Fig. 2** Stribeck curves as a function of the sliding velocity. The data shown was obtained with different buffer solutions on a steel-PDMS material pairing: (a) 10 mM phosphate, 25 mM phosphate, 50 mM phosphate, 100 mM phosphate, 250 mM phosphate, 500 mM phosphate and 20 mM HEPES. (b) UB and BRB at different pH levels (from pH 2 to 8). Error bars denote the standard deviation as obtained from $n$ = 3 different measurements.

A material incompatibility between the friction partners and the buffer substance is not the only issue that needs to be considered when selecting a buffer system. Proteins such as mucins might be able to operate in a similarly wide range of pH settings as they also occur in the human body [42]. In potential applications as a biolubricant, the relevant pH range a mucin solution might encounter can span from slightly alkaline (pH ~ 8 in the oral cavity [43]) to considerably acidic (pH ~ 4 in the vaginal tract [44]).

Hence, conducting measurements at different pH levels is important to assess the working range of mucin solutions. However, only few buffers are efficient over such a wide pH range, and using different buffer substances at different pH levels may lead to artefacts that again can complicate the interpretation of the results. We here compare two established buffer systems which have a broad operating range, i.e., a universal buffer (UB, pH range: 2.0 - 8.2) and a Britton-Robinson buffer (BRB, pH range: 1.9 - 11.0). As presented in **Fig. 2b**, the Stribeck curves obtained with either buffer show a slight pH dependence, and this effect is more pronounced for BRB. Moreover, in the boundary lubrication regime, the friction coefficients measured with BRB are almost one order of magnitude lower than those determined with UB. At this point, it is important to realize that one of the components of BRB is phosphate – which explains the lower friction coefficients compared to UB. Thus, although BRB can operate in a broad pH range, it is not an ideal buffer system for our tribological experiments.

Based on those results, we conclude that phosphate-free buffers are better suited for tribological investigations employing a steel/PDMS paring. Accordingly, for our study, we choose HEPES buffer for experiments at neutral pH, and we use UB in those experiments, where the pH is varied.

## 3.2 Viscosities of the different mucin-based lubricants

Before we study the tribological performance of mucin solutions at different conditions, we first assess the viscosities of the different mucin-based mixtures. These values are summarized in **Table 1** and sorted into groups according to the different parameters (mucin concentration, buffer pH, buffer ionic strength, influence of other proteins) whose influence on mucin lubricity is tested later. As can be expected, increasing the mucin concentration increases the viscosity of the solution, with the 1 % (w/v) mucin solution showing the highest viscosity of 8.3 mPa·s. The viscosities of mucin/protein mixtures are about twice as high as those of solutions containing mucin only – owing to the increased total protein concentration. All other solutions have viscosities around 1 mPa·s similar to pure water [45]. As a consequence, with the exception of the 1 % (w/v) mucin solution, we do not expect that those minor changes in lubricant viscosity will have a perceivable impact on the results obtained with tribology: This is why we do not recalculate the obtained friction data into representations using the Sommerfeld number as a variable, but show the dependencies of the friction factor on the sliding speed in all diagrams.

**Table 1**. Viscosities of different mucin-based solutions. The error values shown depict the standard deviation as obtained from $n = 3$ independent measurements.

| No. | Buffer | pH | Components | Viscosity (mPa·s) |
|---|---|---|---|---|
| A1 | HEPES | 7.0 | 0.005 % mucin | 0.99 ± 0.01 |
| A2 | HEPES | 7.0 | 0.01 % mucin | 0.99 ± 0.01 |
| A3 | HEPES | 7.0 | 0.05 % mucin | 1.07 ± 0.01 |
| A4 | HEPES | 7.0 | 0.1 % mucin | 1.24 ± 0.18 |
| A5 | HEPES | 7.0 | 1.0 % mucin | 8.30 ± 0.05 |
| B1 | UB | 2.0 | 0.1 % mucin | 1.04 ± 0.10 |
| B2 | UB | 4.0 | 0.1 % mucin | 0.92 ± 0.11 |
| B3 | UB | 6.0 | 0.1 % mucin | 1.13 ± 0.42 |
| B4 | UB | 8.0 | 0.1 % mucin | 1.23 ± 0.05 |
| C1 | HEPES | 7.0 | 0.1 % mucin + 20 mM NaCl | 1.22 ± 0.05 |
| C2 | HEPES | 7.0 | 0.1 % mucin + 50 mM NaCl | 1.38 ± 0.11 |
| C3 | HEPES | 7.0 | 0.1 % mucin + 150 mM NaCl | 1.23 ± 0.07 |
| C4 | HEPES | 7.0 | 0.1 % mucin + 500 mM NaCl | 1.19 ± 0.01 |
| D1 | HEPES | 7.0 | 0.1 % mucin + BSA (40 mg/mL) | 2.41 ± 0.49 |
| D2 | HEPES | 7.0 | 0.1 % mucin + Amylase (150 U/mL) | 1.89 ± 0.55 |
| D3 | HEPES | 7.0 | 0.1 % mucin + Lysozyme (13 mg/mL) | 1.84 ± 0.28 |

## 3.3 Influence of the mucin concentration

In a next step, we ask how strongly the mucin concentration affects the lubricity of mucin solutions. This question is motivated by two aspects: first, there might be an optimal mucin concentration above/below which the lubricant is less efficient; second, for economic reasons, identifying the minimal mucin concentration that yields good lubricity is important since the current purification process only produces limited amounts of mucins. We first examine a standard mucin concentration of 0.1 % (w/v) as also used in earlier studies, and we obtain virtually identical results as described before: the mucin solution reduces the friction factor $\mu$ by approximately two orders of magnitude, and this also holds true in the boundary lubrication regime, where we measure $\mu \sim 0.01$ (**Fig. 3a**) – this value is much lower than previous results obtained with commercial mucins [25]. Interestingly, we do not detect any perceivable improvement in lubricity when the mucin concentration is increased to 1.0 % (w/v). At this higher mucin concentration, the viscosity of the solution is already increased 8fold. Thus – at least according to Stribeck theory – an increased lubricant viscosity should improve the lubricity of the system; however, this is not the case, which indicates that lubricant viscosity is not a dominant factor for the tribological performance of mucin solutions. It also suggests that further increasing the mucin concentration will not be of much avail. Instead, we lower the mucin concentration by a factor of ten and find that, even at this reduced mucin content, the lubricity of the solution remains almost unaffected. However, this concentration of 0.01 % (w/v) seems to constitute a threshold: If the mucin solution is further diluted, i.e., to 0.005 % (w/v), the obtained Stribeck curve closely resembles the one determine with simple HEPES buffer (devoid of mucins). This suggests that, at this low mucin concentration, some critical property of the solution is changed such that the lubricity of the system is suddenly and strongly compromised.

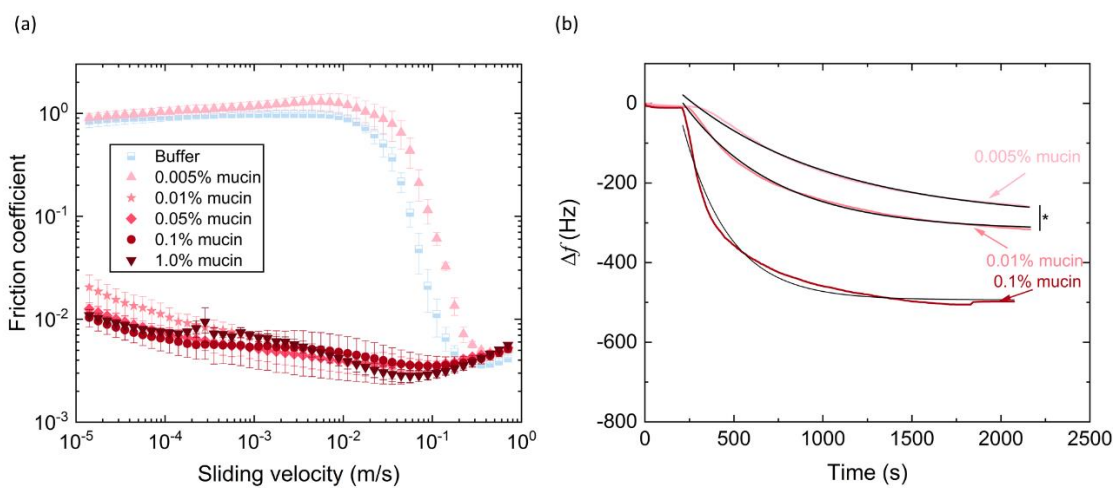

**Fig. 3** Influence of the mucin concentration on the lubricity and adsorption behavior of mucin solutions. (a) Tribological tests were carried out with a steel-PDMS pairing using 20 mM HEPES buffer at pH 7 with mucin concentrations ranging from 0 (pure HEPES buffer) to 1.0 % (w/v). The error bars denote the standard deviation as obtained from $n = 3$ independent measurements. (b) Adsorption behavior of mucin solutions containing different concentrations of MUC5AC as assessed by QCM-D using PDMS-coated Au-chips. The results denote the average as obtained from $n = 5$ ($n = 3$ in case of the 0.1 % mucin solution) independent measurements. Black lines denote the fit curves of the obtained results. The asterisk in (b) indicates statistical significance using a *p*-value of 0.05.

For macromolecular lubricants, the adsorption properties of the solubilized (bio)polymers constitute such a parameter that is closely related to the lubricity of the solution [19, 46]. Accordingly, we test the kinetics of mucin adsorption onto PDMS at different mucin concentrations using QCM-D. As displayed in **Fig. 3b**, at low mucin concentrations of 0.005 % (w/v), we not only detect a significant, ~2-fold reduction (compared to standard concentrations of 0.1 %) in the plateau value of the frequency shift but also slower adsorption kinetics. We quantify this impression by fitting an exponential decay to the QCM-D data: $\Delta f \sim \exp(-t/\tau)$. These fits (**Fig. 3b**) demonstrate that this decay time $\tau$ changes from ~250 s (0.1 % mucin) to ~500 s (0.01 % mucin), and even further to ~1000 s (0.005 % mucin). Since, in the last mucin dilution step, the adsorption kinetics seem to change more critically than the plateau value of $\Delta f$, we speculate that the speed of mucin adsorption limits mucin lubricity at low concentrations. In a model where sacrificial layer formation and hydration lubrication are the two key mechanisms governing this lubricity, this can be rationalized very well: due to the tribological stress acting on the PDMS surface, adsorbed mucins are constantly sheared-off and can only provide hydration lubrication when they readsorb quickly enough. If this readsorption process is too slow, e.g., when the amount of available mucin macromolecules in the lubricant solution is below a sub-critical concentration, both sacrificial layer formation and hydration lubrication are compromised at the same time – resulting in drastically reduced lubricity as observed here.

## 3.4 Influence of pH

When used as a biomedical lubricant, reconstituted mucin solutions might encounter a broad range of pH levels in the human body. Thus, we next evaluate the lubrication and adsorption properties of mucin solutions at different pH levels. As shown in **Fig. 4a**, when the pH value of a mucin solution is slightly increased from 7 to 8, the measured friction coefficients are slightly higher. The opposite effect is observed in the acidic regime; here, lowering the pH to 4 moderately improves the lubricity of the mucin solution, and we obtain friction coefficients smaller than 0.01 over the whole range of sliding speeds tested. However, further decreasing the pH does not continue this trend. At strongly acidic levels (i.e., at pH 2), the measured friction curve again resembles that obtained at neutral pH. In other words, as indicated in the inset of **Fig. 4a**, there seems to be an optimal pH regime where the mucin solution performs best. This pH regime coincides with the pH of the vaginal fluid outside ovulation [44], indicating that a reconstituted mucin solution could offer supreme lubricity when applied as a personal lubricant for intercourse. Interestingly, MUC5AC as investigated here not only occurs in the gastric mucosa (from which it was purified) but also in cervicovaginal mucus. Other mucin variants obtained from different sources might have different pH optima regarding lubricity. However, even though we find that mucin lubricity is best at pH 4, the friction coefficients obtained at the other physiologically relevant pH levels are also on the order of 0.01, which is still excellent regarding potential biomedical applications [47, 48].

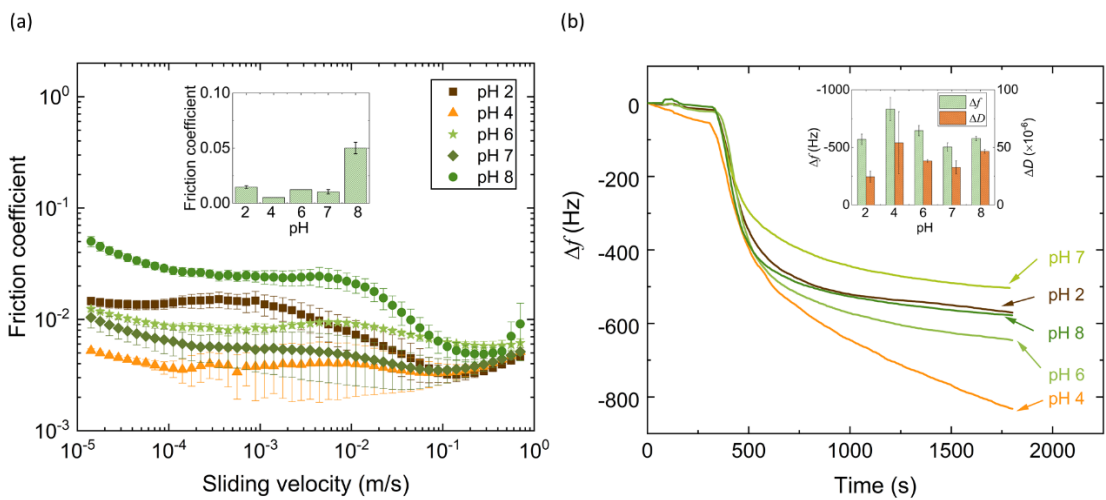

**Fig. 4** Impact of pH on the lubricity and adsorption properties of 0.1% (w/v) mucin solutions solubilized in UB buffer. (a) Tribological measurements were conducted with a steel-PDMS paring tuning the pH from 2 to 8. The error bars denote the standard deviation as obtained from $n = 3$ independent measurements. (b) Adsorption properties (plateau shift in resonance frequency $\Delta f$ and plateau dissipation shift $\Delta D$ – full $\Delta D$ curves are shown in Fig. S3) of 0.1 % (w/v) mucins onto PDMS-coated Au-chips at different pH levels. The error bars in the inset denote the standard deviation as obtained from $n = 3$ independent measurements.

Having established that alterations in pH do affect the lubricity of mucin solutions, we now ask why this is the case. Also here, we assess the adsorption behavior of mucins with QCM-D. This time, as shown in **Fig. 4b**, there is no major change in the adsorption kinetics – which agrees with our finding that mucin lubricity is very good at all pH levels tested. However, we find that the final frequency shift reaches its lowest level at pH 4 both onto PDMS and steel surfaces (see inset of **Fig. 4b and Fig. S2a**). This indicates that, here, the adsorbed mucin layer assumes an optimal configuration. Similarly, also the dissipation shift Δ*D* exhibits a local extremum at pH 4 (inset of **Fig. 4b and Fig. S3**), which underscores our assumption that some transition in the properties of adsorbed mucins occurs here. It was suggested that the conformation of gastric mucin changes at this pH[49], and we also detect differences in the hydrodynamic size of mucins when probed with dynamic light scattering at different pH levels (**Fig. S4**). It was suggested before that this alteration in mucin structure also impacts the rheological properties of mucin solutions[50]. For our data, there are two different explanation alternatives as to why the measured frequency and dispassion shifts are strongest at pH 4 (**Fig. S3**): First, it is possible that more mucin molecules adsorb at this pH level than at higher/lower pH because, here, the mucin configuration is optimal for promoting mucin adsorption onto PDMS – and this would improve the efficiency of sacrificial layer formation. Second, the adsorbed mucin layer might be slightly better hydrated at pH 4 thus improving its hydration lubrication capabilities.

## 3.5 Influence of the NaCl concentration

The next physiological parameter we explore regarding its influence on the lubricity of mucin solutions is the NaCl concentration. NaCl plays a vital role in the regulation of many body functions and is also an important part of the body's fluid balance control system. However, the NaCl concentration in those body fluids can change, e.g. as a result of electrolyte disturbance [51]. The same holds true for the ion concentration in mucosal layers, which can be affected by disease or after the consumption of food or beverages.

As displayed in **Fig. 5**a, the lubricity of solutions comprising manually purified mucin is indeed sensitive to the NaCl concentration in the mucin buffer. Again, this is different from previous results obtained with commercial mucins [52]. When the NaCl content in a mucin solution is increased from low (i.e., 20 mM or 50 mM) concentrations to 150 mM, a slight increase of the friction coefficient is observed in the boundary and mixed lubrication regime – but not in the hydrodynamic regime. The latter agrees with our previous observation that – at mucin concentrations of 0.1 % (w/v) – increasing the ionic strength of the solution does not affect its viscosity. This effect becomes a bit more pronounced at high NaCl concentrations (i.e., 500 mM); however, also here, we still measure friction coefficients well below 0.1 at all sliding velocities tested. This shows that the lubricity of mucin solutions is also quite robust towards variations in this physiological parameter. Moreover, it indicates that purified mucins may operate well as lubricants on various body surfaces even if they exhibit large differences in ionic strength.

Even though the lubrication potential of mucin solutions is not strongly reduced by increasing salt concentrations, we ask by which mechanism mucin lubricity is affected by NaCl. Since mucins adsorb very well to PDMS by means of hydrophobic interactions [19], it appears unlikely that NaCl can interfere with this process. Indeed, QCM-D measurements demonstrate that the adsorption kinetics of mucins onto PDMS do not change much with increasing NaCl concentrations (**Fig. 5b**). However, the plateau values in the resonance frequency shift $\Delta f$ become larger at higher NaCl contents, both on PDMS-coated Au-chips (**Fig. 5b**) and steel-chips (**Fig. S2b)**. This suggests that neither the speed of mucin adsorption nor the amount of adsorbed mucins in equilibrium are responsible for the observed reduction in lubricity. Thus, we further investigate the physical properties of the adsorbed mucin layers and evaluate alterations in the dissipation shift $\Delta D$ – another parameter that can be obtained by QCM-D. As depicted in **Fig. 5b**, this parameter is changed with the same kinetics as the frequency shift $\Delta f$, but increases as $\Delta f$ decreases. Higher NaCl concentrations induce a stronger alteration in both, $\Delta D$ and $\Delta f$, and sequential exposure of PDMS-coated QCM-chips to mucin and NaCl solutions returns a similar picture. The latter illustrates that NaCl can also modify the properties of pre-adsorbed mucins – at least at high enough NaCl concentrations (**Fig. 5c**).

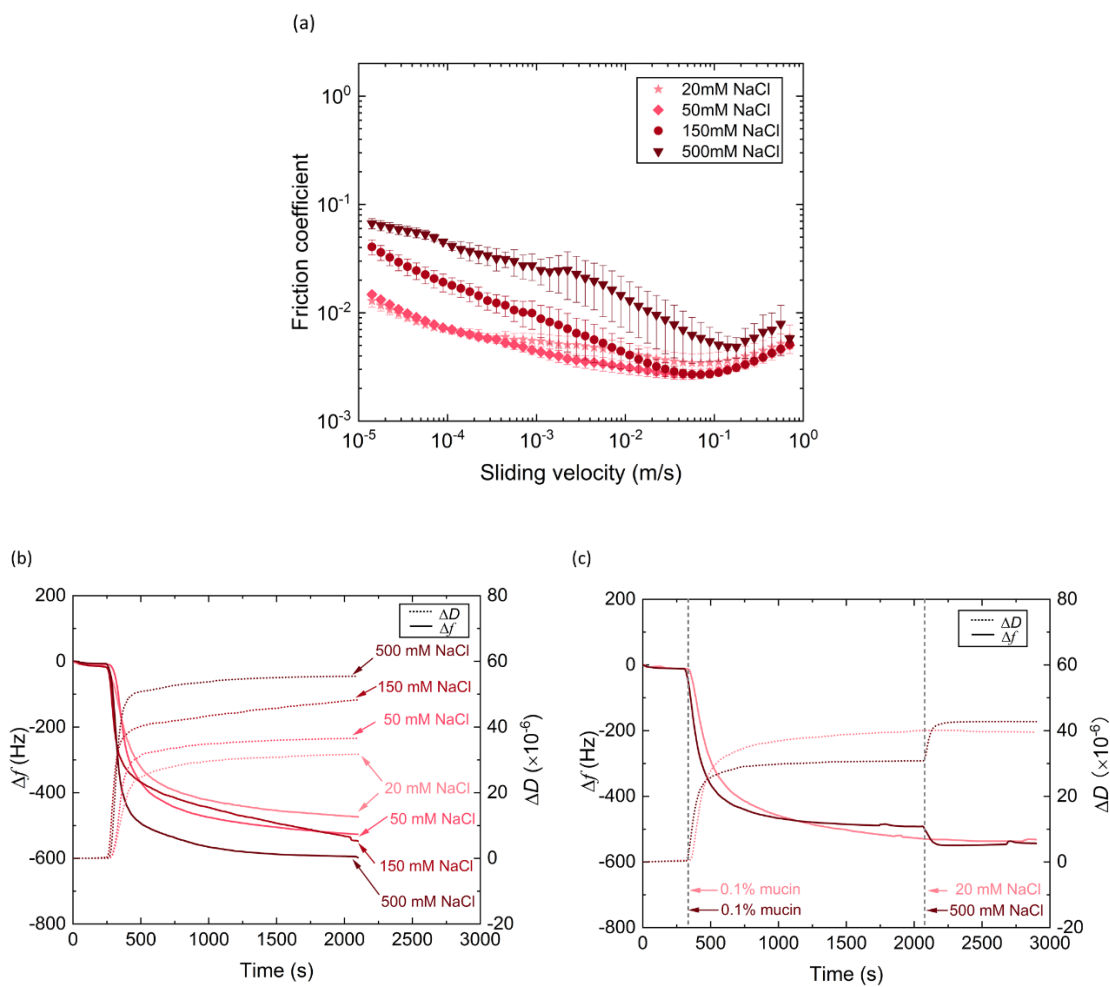

**Fig. 5** Impact of the NaCl concentration on the lubricity and adsorption properties of 0.1% (w/v) mucin solutions. (a) Tribological measurements were conducted with a steel-PDMS pairing using different NaCl concentrations as indicated in the graph. Error bars denote the standard deviation as obtained from $n = 3$ different measurements. (b) Changes in frequency shift (solid line) and dissipation shift (dashed line) for PDMS-coated Au-chips when brought in contact with mucin solutions containing different concentrations of NaCl. (c) Sequential exposure of PDMS-coated Au-chips to a 0.1% (w/v) mucin solution followed by rinsing with either a 20 mM or a 500 mM NaCl solution. The curves shown denote averages as obtained from $n = 3$ independent measurements.

There are several possibilities to rationalize this observation: High ion concentrations induce Debye screening and thus reduce the number of charged groups available for ionic paring [53]. In addition, it has been suggested that a shortened Debye-length can decrease the range and magnitude of the steric repulsion forces acting between mucin layers [54]. In salt solutions, mucins are reported to undergo a conformational change from a fully extended state to a collapsed state, and – consistently, we observe a decrease of the hydrodynamic size of mucins in the presence of NaCl when probed with dynamic light scattering (**Fig. S4**). This behavior was suggested to originate from a combination of two effects: electrostatic screening and osmotic pressure acting on the mucin chains [55]. Together, those effects can influence the structure of an adsorbed mucin layer, thus increasing its stiffness and/or reducing its water content. Dehydration of mucin results in gradual collapse of the mucin chains and a decrease of the compatibility between mucin and an aqueous solution [56]. This might limit the hydration lubrication abilities of a surface bound mucin layer, which would explain the higher friction coefficients observed in our tribological experiments.

## 3.6 Influence of proteins

When applied to mucosal surfaces of the human body, mucin solutions will not only encounter pre-existing pH conditions and NaCl concentrations, but also a variety of proteins. Thus, in a last step, we ask if and how such proteins might interfere with the lubricity and mucin solutions. Lysozyme, amylase, and serum albumin are prominent examples of proteins occurring in body fluids such as tears, saliva, synovial fluid and vaginal fluid [57-60], and we now assess the influence of each of these proteins independently. The friction curves obtained for 0.1 % (w/v) mucin solutions determined in the presence of one of those proteins are shown in **Fig. 6a**. For comparison, also results obtained with a pure mucin solution and simple HEPES buffer are shown.

For all protein/mucin mixtures tested, higher friction coefficients are obtained compared to when a pure mucin solution is used for lubrication. If BSA or amylase is added to the mucin solution, the friction coefficients at very low sliding speeds are strongly increased and reach values almost as high as those obtained with simple HEPES buffer. For lysozyme, this effect is less drastic, and we only detect an increase in friction by approximately an order of magnitude. In the mixed lubrication regime, i.e., at sliding velocities in the range of $10^{-3}$ to $10^{-1}$ m/s, we find a noticeable friction reduction compared to simple buffer in all cases. Here, all protein/mucin mixtures return friction coefficients that are in between those obtained with pure mucin solutions and pure buffer. Finally, in the hydrodynamic lubrication regime, all friction curves overlay, which fully agrees with our finding that the viscosities of the protein/mucin mixtures are similar to that of pure mucin solutions.

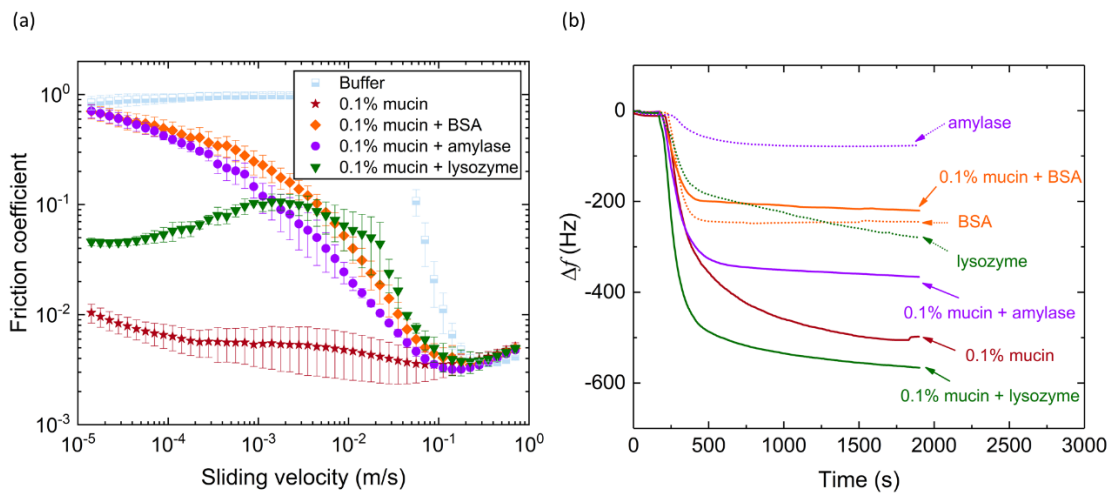

**Fig. 6** Lubrication and adsorption properties of mucin solutions in the presence of selected proteins. (a) Tribological tests were conducted with a PDMS/steel pairing lubricated with 0.1 % (w/v) mucin solutions both in the presence and absence of BSA, amylase and lysozyme, respectively. The blue squares denote results obtained with simple HEPES buffer. Error bars denote the standard deviation as obtained from $n = 3$ different measurements. (b) QCM-D measurements depict alterations in the frequency shift ($\Delta f$) as a function of time when probing the adsorption of BSA, amylase or lysozyme (in 20 mM HEPES buffer, pH 7) onto PDMS-coated Au-chips in the presence (solid line) or absence (dashed line) of 0.1 % (w/v) mucin, respectively. The curves shown denote the average as obtained from $n = 3$ independent measurements.

One possible explanation of how these proteins might interfere with the lubricity of mucin solutions could be that the proteins negatively affect the adsorption behavior of mucins. We test this hypothesis by conducting QCM-D measurements with the three mucin/protein mixtures and compare the results to the adsorption behavior obtained with the same protein solutions in the absence of mucins. As depicted in **Fig. 6b**, the combined adsorption of mucin and BSA shows almost no difference compared to the adsorption kinetics of a pure BSA solution alone. This indicates, that BSA blocks the surface for the adsorption of mucin – which is consistent with the regular use of BSA as a blocking agent in biochemical lab assays. Thus, in this particular case – as a direct consequence of suppressed mucin adsorption – both key mechanisms required for mucin lubricity, i.e. sacrificial layer formation and hydration lubrication, are compromised, which leads to highly increased friction coefficients.

For the other two combinations, i.e., mucin/amylase and mucin/lysozyme, we observe a stronger frequency shift compared to the when the corresponding small protein is tested alone. This shows that, different from what we described above for BSA, in the presence of either lysozyme or amylase, mucin adsorption is still possible – either onto free areas not covered by the smaller proteins or onto pre-adsorbed lysozyme/amylase layers. However, the lubricity of the mucin/lysozyme solution is similarly compromised as we observed for mucin/BSA, whereas the lubricity of the mucin/lysozyme combination is still decent – especially in the boundary lubrication regime (**Fig. 6a**).

The molecular weight of BSA, amylase, and lysozyme is ~66 kDa, ~51 kDa and ~14 kDa, respectively – all these proteins are much smaller than mucin (0.5-20 MDa – depending on the degree of oligomerization [61]). Accordingly, the smaller protein molecules can diffuse through the aqueous solution more rapidly and might adsorb onto the PDMS surface more quickly than mucin, and mucins could adsorb to this pre-adsorbed layer of small proteins. Indeed, sequential exposure to the small proteins followed by mucin adsorption underscores this notion (**Fig. S5**). Then, the presence of this protein pre-coating could influence mucin adsorption differently – depending on the properties of this pre-coating layer. At neutral pH, which is studied in this scenario, the net charge of mucin, amylase is negative whereas lysozyme is positively charged [62, 63]. Thus, now the mucins would now adsorb to the pre-coated PDMS surface by means of electrostatic forces instead of hydrophobic forces. We speculate, that such an electrostatically bound mucin layer might be less favorable for providing lubricity by means of sacrificial layer formation, and that the binding strength of the mucin (which is overall polyanionic but carries both negatively charged and positively charged motifs) to cationic and anionic surfaces is different. How such a difference in mucin binding strength to the surface it has adsorbed to would affect the lubricity of the system is, however, not clear at this point.

# 4 Conclusions

In this work, we have evaluated the interplay of viscosity, adsorption kinetics, and lubricity of solutions comprising manually purified porcine gastric mucins on hydrophobic surfaces. Different pH, proteins, mucin and salt concentrations were investigated to mimic different physiologically relevant conditions. We here demonstrate that mucin solutions provide superior lubricity even at very low concentrations of 0.01 % (w/v), over a broad range of pH levels and even at elevated ionic strength. Moreover, we show that different proteins such as amylase, BSA, or lysozyme reduce the lubricity of mucins solutions – especially in the boundary lubrication regime. In the mixed lubrication regime, where physiological processes such as intercourse, blinking, swallowing or chewing can be expected to occur, this limitation is less pronounced and we still measure friction coefficients which are well suitable for mucin solutions to be applied as biomedical lubricants [47, 48].

We here have used a combination of artificial materials to probe the lubricity of mucin solutions and to elucidate microscopic mechanisms which regulate mucin lubricity. The results imply that manually purified mucin solutions are very robust towards alterations in the lubricant composition as they might occur when a mucin solution is applied to a mucosal tissue. Our data suggests that mucin solutions can be powerful lubricants for several biomedical applications. Certainly, all of those applications will come with their own, detailed challenges in using mucin solutions as lubricants; those specific challenges can stem from differences in how the mucin solution is applied (e.g., as a spray for oral lubrication vs. drops for ocular lubrication) or other, more biologically-driven issues (e.g., interaction of mucins with food particles when used as a mouth spray or with foreign objects such as contact lenses). However, the physiological parameters we study in this manuscript apply to all those applications targeting (human) mucosal tissues as existing conditions on those tissues (pH, NaCl concentration, presence of small proteins) will modify the lubrication properties of mucin solutions right after their application.

## Conflicts of interest

There are no conflicts of interest to declare.

## Acknowledgments

We thank Matthias Marczynski for assistance with the mucin purification. This project has received funding from the European Union's Horizon 2020 research and innovation programme under the Marie Skłodowska-Curie grant agreement No. 754462.

## Supporting Information

A PDF file with data on the lubricity of mucin solutions tested with different material pairings, the adsorption properties of mucins onto steel-chips and PDMS-coated Au-chips, the hydrodynamic size of mucins at different conditions, and sequential adsorption measurements of proteins and mucins onto PDMS-coated Au-chips is available as supplement.

*Volume two.* **1956,** (4th edition).

25. Lee, S.; Müller, M.; Rezwan, K.; Spencer, N. D., Porcine gastric mucin (PGM) at the water/poly (dimethylsiloxane)(PDMS) interface: influence of pH and ionic strength on its conformation, adsorption, and aqueous lubrication properties. *Langmuir* **2005,** *21* (18), 8344-8353.

26. Martins, A. M.; Santos, M. I.; Azevedo, H. S.; Malafaya, P. B.; Reis, R. L., Natural origin scaffolds with in situ pore forming capability for bone tissue engineering applications. *Acta Biomaterialia* **2008,** *4* (6), 1637-1645.

27. Rettig, R.; Virtanen, S., Composition of corrosion layers on a magnesium rare‐earth alloy in simulated body fluids. *Journal of Biomedical Materials Research Part A: An Official Journal of The Society for Biomaterials, The Japanese Society for Biomaterials, and The Australian Society for Biomaterials and the Korean Society for Biomaterials* **2009,** *88* (2), 359-369.

28. Boettcher, K.; Grumbein, S.; Winkler, U.; Nachtsheim, J.; Lieleg, O., Adapting a commercial shear rheometer for applications in cartilage research. *Review of Scientific Instruments* **2014,** *85* (9), 093903.

29. Goda, T.; Konno, T.; Takai, M.; Moro, T.; Ishihara, K., Biomimetic phosphorylcholine polymer grafting from polydimethylsiloxane surface using photo-induced polymerization. *Biomaterials* **2006,** *27* (30), 5151-5160.

30. Hanawa, T., In vivo metallic biomaterials and surface modification. *Materials Science and Engineering: A* **1999,** *267* (2), 260-266.

31. Placette, M. D.; Himes, A. K.; Schwartz, C. J., Investigation of Wear Mechanisms in Silicone Sleeved Implantable Cardiac Device Leads using an In Vitro Approach. *Biotribology* **2019,** *17*, 40-48.

32. Mills, T.; Norton, I. T.; Bakalis, S., Development of tribology equipment to study dynamic processes. *Journal of Food Engineering* **2013,** *114* (3), 384-390.

33. Çelebioğlu, H. Y.; Gudjónsdóttir, M.; Chronakis, I. S.; Lee, S., Investigation of the interaction between mucins and β-lactoglobulin under tribological stress. *Food Hydrocolloids* **2016,** *54*, 57-65.

34. Ghosh, S.; Choudhury, D.; Roy, T.; Moradi, A.; Masjuki, H. H.; Pingguan-Murphy, B., Tribological performance of the biological components of synovial fluid in artificial joint implants. *Science and technology of advanced materials* **2015,** *16* (4), 045002.

35. Yakubov, G. E.; Macakova, L.; Wilson, S.; Windust, J. H. C.; Stokes, J. R., Aqueous lubrication by fractionated salivary proteins: Synergistic interaction of mucin polymer brush with low molecular weight macromolecules. *Tribology International* **2015,** *89*, 34-45.

36. Menezes, P. L.; Kailas, S. V.; Lovell, M. R., Friction and transfer layer formation in polymer–steel tribo-system: role of surface texture and roughness parameters. *Wear* **2011,** *271* (9-10), 2213-2221.

37. Myant, C.; Spikes, H. A.; Stokes, J. R., Influence of load and elastic properties on the rolling and sliding friction of lubricated compliant contacts. *Tribology International* **2010,** *43* (1-2), 55-63.

38. Li, J.; Zhou, F.; Wang, X., Modify the friction between steel ball and PDMS disk under water lubrication by surface texturing. *Meccanica* **2011,** *46* (3), 499-507.

39. Winkeljann, B.; Bussmann, A. B.; Bauer, M. G.; Lieleg, O., Oscillatory Tribology Performed With a Commercial Shear Rheometer. *Biotribology* **2018,** *14*, 11-18.

40. Zhang, L.; Feng, D.; Xu, B., Tribological Characteristics of Alkylimidazolium Diethyl Phosphates Ionic Liquids as Lubricants for Steel–Steel Contact. *Tribology Letters* **2009,** *34* (2), 95-101.

# Table of Contents

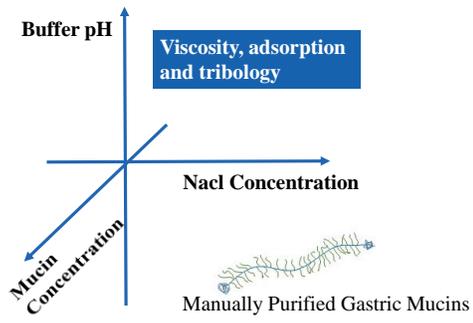

Manually Purified Gastric Mucins